# Comparison of structure making/breaking properties of alkali metal ions $Na^+$, $K^+$ and $Cs^+$ in water

Sudakshina Roy[1] and Barnana Pal[*]

Saha Institute of nuclear Physics,
Condensed Matter Physics Division,
1/AF, Bidhannagar, Kolkata ‑ 700064, India

Abstract: The alkali metal ions in aqueous electrolyte solutions have strong influence on the surrounding network structure of water formed through hydrogen bonds. The extent of ionic perturbation to the structure of water depends on the nature of individual ions and the solute concentration. Experimental techniques like Neutron diffraction, X-ray diffraction, Raman spectroscopy, isotopic substitution study, viscosity measurements and related theoretical or simulation studies are used to understand the characteristics features of the structural changes in these solutions. Recent ultrasonic studies on aqueous solutions of NaCl, KCl and CsCl show anomalous changes in the variation of velocity (v) with increase in concentration (c) from very dilute to saturation limit. The experimental observations and theoretical or simulation studies available in the literature are considered to make a comparative study on the structure making/breaking properties of $Na^+$, $K^+$ and $Cs^+$ ions in water and consequent effect in ultrasonic velocity change.

## 1. Introduction

Study on the structural properties of water has got importance from ancient time due to its essentiality in the formation of life on this Earth. Water, in pure form and as a natural solvent, exhibit interesting and anomalous behavior. An extensive study has been carried out to understand the structure of water [1] and it has been accepted that liquid water consists of a connected network structure of hydrogen bonds (HB). Strained and broken bonds along with unbounded water molecules are embedded in the bulk water and the system is under continuous structural reformation depending upon temperature, pressure and other external parameters.

Addition of ions has strong effect on the HB structure [2,3,4]. An alkaline salt ZA separates into cation ($Z^+$) and anion ($A^-$) in aqueous solutions. The water molecules are dipolar in nature. It has a slightly bent structure with negatively charged oxygen at one end bending towards the two positively charged hydrogen atoms on the other side. The dipolar water molecules get oriented around the charged ions due to electrostatic interaction among them. This is termed as structure making effect. Structure breaking effect is visualized when most or none of the water molecules in a solution are influenced by the presence of ions [5,6]. Modern techniques [4,7,8,] like Raman Spectroscopy, X- Ray diffraction Spectroscopy and Neutron diffraction are employed to explore the underlying mechanism responsible for the ordered structure in water. The water structure making /breaking concept by positive and negative ions was studied [5,6] using Neutron Diffraction data with hydrogen isotope substitution. Collection and interpretation of data obtained from these measurements together with computer simulation method [5,7,9] give useful information about molecular structure of water in presence of solute. Still, there are discrepancies and disagreement between experimental observations as well as simulation data leaving scope of further study on hydrogen bonding and its extent in water and aqueous solutions.

---

[*] Corresponding author
[1] CSIR-Central Glass and Ceramic Research Institute, Kolkata (retired)

The motivation behind our present work is to study the solute-water interaction using the ultrasonic technique, the Fourier Spectrum Pulse-Echo (FSPE) method [10] for the determination of wave propagation parameters through three different alkaline salt solutions NaCl , KCl and CsCl having concentrations from very dilute to saturation limit [11]. The observed anomaly in the variation of velocity (v) with concentration (c) is analyzed in the light of other experimental and simulation results [4,7-9] available in the literature. Section 2 describes most recent results of ultrasonic experiments [11], sec. 3 gives analysis of the data in view of other experimental and simulation data and sec. 4 gives the summary.

## 2. Experimental observations with FSPE

An exploratory work using ultrasonic method, specifically the Fourier Spectrum Pulse-Echo (FSPE) method [10] has been carried out to study the solute-water interaction in three different alkaline salt solutions of NaCl , KCl and CsCl in the concentration range starting from very dilute up to saturation limit [11]. The solutes with different ionic radii r (r = 1.02A° for $Na^+$, 1.38A° for $K^+$ , 1.67A° for $Cs^+$ and 1.84A° for $Cl^-$) are selected to make a comparative study of structure making and breaking effects when dissolved in water. The measurements are done at room temperature ($25^o$ C) at ultrasonic frequencies 1MHz and 2MHz. Here we mention only the anomalous behavior observed in the variation of v with c. Figures 1(a) and (b) show the nature of variations of $v \times 10^{-5}$ (cm/sec) in aqueous solution of NaCl with molar concentration c (mol/L) for two frequencies 1MHz and 2MHz respectively. With the increase of c, v increase. In the concentration range 3-4 mol/L anomalous change in v is observed which is more prominent at 2MHz frequency. Above 4 mol/L, v increases further with increasing c till saturation is reached.

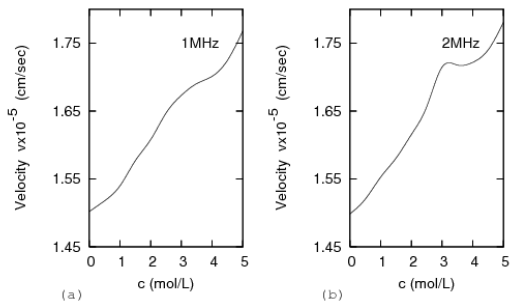

Figure1: Change in velocity $v \times 10^{-5}$ (cm/sec) with concentration c (mol/L) in aqueous NaCl for (a) 1MHz and (b) 2MHz.

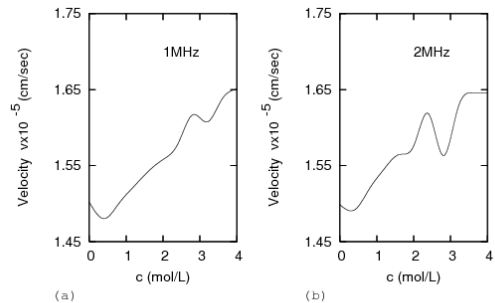

Figure2: Change in velocity $v \times 10^{-5}$ (cm/sec) with concentration c (mol/L) in aqueous KCl for (a) 1MHz and (b) 2MHz.

For aqueous KCl solution the nature is shown in figures 2(a) and (b) for frequency 1MHz and 2 MHz respectively. In this case an initial decrease in v is noticed with increasing c. Then v increases gradually and a similar anomaly as seen for NaCl solution, is observed for c values in the range 2.5-3.5 mol/L. For KCl a prominent discontinuity in v is observed in this concentration range and the effect is more visible for 2 MHz frequency. Above 3.5 mol/L, v values increase further till saturation is reached.

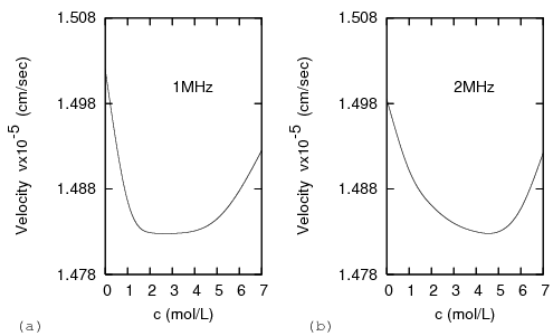

Figure 3: Change in velocity $v \times 10^{-5}$ (cm/sec) with concentration c (mol/L) in aqueous CsCl for (a) 1MHz and (b) 2MHz.

For aqueous CsCl solution, the results are completely different. The variations of v with c for two frequencies 1MHz and 2MHz are shown in Figures 3 (a) and (b) respectively. Figure 3 (a) shows decreasing nature of v with increasing c for c <2 mol/L and then it remains more or less constant up to 4 mol/L. Above 4 mol/L, v increases a little as c is increased up to the saturation level. Figure 3(b) shows similar nature except that the initial decrease in v with increasing c extends almost up to 4 mol/L. Between c values 4-5 mol/L v remains almost constant and then increases a little.

## 3. Data analysis:

The notion of partial pair correlation function p(r), r being the linear distance from the centre of mass of the water molecule was introduced to explain the interaction between the component particles of the water molecules. Information about molecular structure of pure water may be provided by three partial pair correlation functions p(O-O,r), p(H-H,r) & p(O-H,r) [2]. To understand the interaction in presence of the solutes in water an additional number of partial correlation functions have been introduced by other researcher [5,6]. These functions are termed as p(Z-O,r), p(Z-H,r), p(A-O,r) and p(A-H,r) describing the ion-water interaction and p(Z-A,r), p(Z-Z,r) & p(A-A,r) describing the ion-ion interaction.

The nature of the three functions p(O-O,r), p(H-H,r) & p(O-H,r) representing pure water structure were studied with various techniques like X-ray diffraction, Raman spectroscopy and neutron diffraction and simulation with structural modeling [2,5,6,8]. The function p(O-O,r) is responsible for structure breaking/making of water. Figure 4 shows the typical double peak structure observed for pure water in the Fourier transform F(Q) of the partial structure factor p(O-O,r) [6]. The neighboring second peak is a measure of disruption degree of the hydrogen bond. The other two functions p(O-H, r) and p(H-H, r) do not have any significant variation over the concentration range [5]. The contributions of these functions are small due to small atomic number of hydrogen [9]. The HB length decrease continuously with increasing solvent concentration and the nature of the double peak gets narrower as recorded in X-ray diffraction and Raman spectroscopy study [8]. The second peak shifts towards left indicating more and more breakdown of the HB bonds with increasing concentration. Therefore we can say that all of the three ions $Na^+$, $K^+$ and $Cs^+$ ions have their contributions to perturb the water structure.

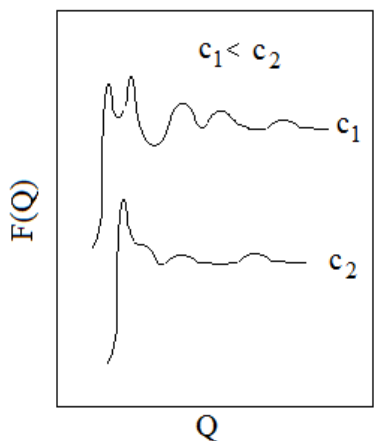

Figure 4: Nature of composite partial structure factor F(Q) for two different conc. $c_1$ and $c_2$, $c_1 < c_2$.

Among the four functions p(Z-O,r), p(Z-H,r) p(A-O,r) p(A-H,r) corresponding to water-ion interaction, p(Z-O,r) and p(A-O,r) change considerably with increasing c for all the three solvents taking part in the structure breaking /making process. Typical nature of partial pair correlation functions p(X-O,r) is shown in figure 5, where X represent any of the ions $Na^+$, $K^+$, $Cs^+$ and $Cl^-$. Comparison of the functions p($Na^+$-O,r), p($K^+$-O,r) p($Cs^+$-O,r) and p($Cl^-$-O,r) [5] reveals following interesting features. The function p($Na^+$-O,r) shows steeper 1st and 2nd maxima compared to the p($K^+$-O,r), p($Cs^+$-O,r) and p($Cl^-$-O,r) functions due to small sizes of $Na^+$ ions. The peak heights change with increase in salt concentration for all the solutions and the 2nd maxima shifts towards left showing change in the water structure by all the ions. The lower peak intensity proves the water structure breaking tendency of $Cs^+$, $K^+$ ions compared to $Na^+$ ions. The p($Na^+$-H,r), p($K^+$-H,r), p($Cs^+$-H,r) and p($Cl^-$-H,r) functions do not vary appreciably in case of pure water structure.

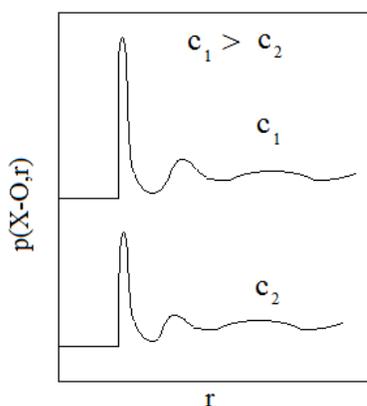

Figure 5: Nature of partial pair correlation function p(X-O,r) at two different concentrations $c_1$ and $c_2$, $c_1 > c_2$. X represents corresponding cation or anion.

The nature of the functions p(A-O,r), p(A-H,r) are almost same at low salt concentrations. Minor differences in peak intensities and positions of the 1st and 2nd maxima show small rise as c increases. This indicates that perturbation to the water structure due to the presence of Cl⁻ is not significant. Only one hydrogen atom of each water molecule points towards the Cl⁻ leaving the remaining hydrogen atom available for bonding to other water molecules or atoms.

The schematic diagram of a water molecule and possible orientations of these molecules around positive and negative ions in solutions are shown in fig 6(a), (b) and (c) respectively. It is obvious that water molecules orient with the oxygen end towards the positive ions as shown in fig (b) and around the negative ions the orientations will be like that represented in fig (c).

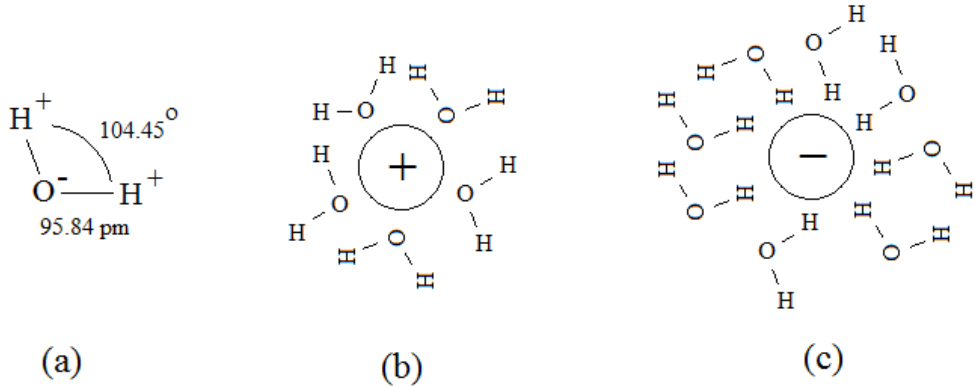

Fig 6. Schematic diagram of a water molecule (a) and possible orientations of water molecules around positive (b) and negative (c) ions.

Figure 7 shows a comparative representation of the angular distribution functions $P(\theta_{ZOH})$ of the angle formed between Z-O or Z-H director and water dipole moment [6] for NaCl and KCl. The curve for NaCl shows finer and steeper peak. For KCL solution the angular distribution peak is wider implying that the water molecules show better ordering for NaCl solution than in KCl solution at higher concentration range.

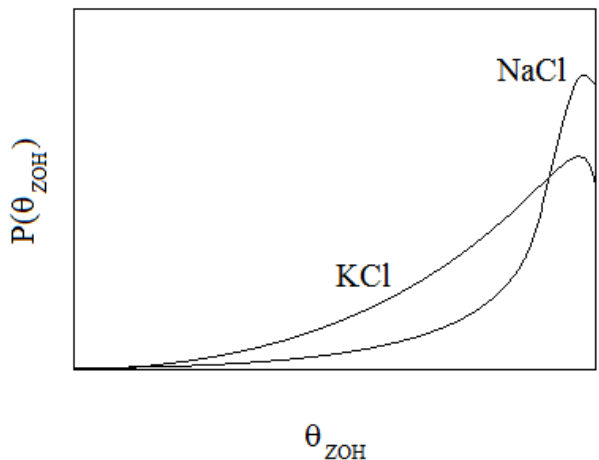

Figure 7. Distribution function of the angle formed between the cation-water oxygen director and the dipole of water molecules within the cation hydration shell in solution of NaCl and KCl.

The other functions p(Z-A,r), p(Z-Z,r) and p(A-A,r) describing the ion-ion correlations play important role particularly at high solute concentration range when ion pairing may occur [5].The ion -ion interaction depends on the solubility, mobility and ionic radii of the solutes. Among the three salts the solubility is highest for NaCl and lowest for CsCl. From the radial distribution functions obtained in Neutron Diffraction experiments [5] and molecular dynamics simulation [6] it is observed that cation-cation interaction is highest in NaCl solution and lowest in CsCl. On the other hand anion-anion interaction becomes more prominent in case of NaCl solution at high concentration. The peaks observed in the function p(Z-Z ,r) is more prominent in case of $Cs^+$ ion implying a strong structure breaking effect for water [5]. Overlapping peaks are observed in the functions p(Z-A,r) ,p(Z-O,r) , p(A-O,r) p(O-O,r) for each of the three solutions indicating no preferred ordered structure in the higher concentration region.

From the molecular dynamics simulation [7] it is observed that mobility of $Na^+$ ions is lower than $K^+$ and $Cs^+$. Mobility of $K^+$ is slightly lower than $Cs^+$. Mobility of $Cl^-$ is comparable to $Cs^+$ . Thus from mobility point of view $Na^+$ ions may present favorable condition to reorient the water molecules around it. So NaCl solution may be termed as a strong structure maker. The mobilities of $Cs^+$ and $Cl^-$ are almost same and therefore none of them can be emerged as structure maker. Therefore CsCl is a structure breaking solution. Mobility of $K^+$ is slightly less than $Cl^-$. Thus KCL solution shows a weaker structure making effect than NaCl. The nature of the curves in figs. 1, 2 and 3 may be explained in view of the above experimental findings. The gradual rise of v with c in case of NaCl solution for both frequencies up to a certain level implies its strong structure making property. The nature of curves obtained for KCl solutions for the same concentration range do not show such continuous rise. The curves show an initial negative slope. Then they rise with a slope less steeper than those observed in NaCl solutions indicating that KCl is weaker structure maker compared to NaCl. For CsCl solution the negative slopes continue up to a higher c value, almost up to 5 mol/L, indicating a structure breaking effect for this salt. The rise for c >5 mol/L is due to $Cs^+$-$Cs^+$ ion pairing effect that is evident from the radial distribution function [4].

The solubility of NaCl, KCl and CsCl in water are in decreasing order and the ionic radii of $Na^+$, $K^+$ and $Cs^+$ are in increasing order. The initial negative slope for KCl arises due to its lesser solubility of KCl compared to that for NaCl. The continuous negative slope for CsCl arises due to the least solubility of this salt in water. The typical nature of water structure changes accordingly with increase in c as shown in fig 4 for both NaCl and KCl solutions up to ~3mol/L. The change is more prominent in case of NaCl than KCl and CsCl for increasing c as is evident from the nature of partial pair co relation function shown in fig 5. These results reflect in the nature of the curves shown in figs 1, 2 and 3 asserting NaCl, KCl and CsCl as decreasing water structure breaking properties. The sharper nature of angular distribution functions for NaCl shown in Fig 7 indicates finer orientations of water molecules around $Na^+$ ions compared to that around $K^+$ ions. For $Cs^+$ ions no such structural orientation is found. Isotope effects on transport coefficients and micro dynamical properties indicate structure breaking effect of $Cs^+$ ions in aqueous CsCl solution [12].

The regions in the curves of figs. 1 and 2 showing anomaly in the higher concentration region around 3mol/L may be explained by the ion-ion and ion-water interactions. The radial distribution functions obtained from X-ray diffraction studies [4] show an overlapping region near this concentration value where the ion-water and ion-ion interactions occur simultaneously. In this region no well defined ordering of water molecules can be found. This is indicated by a discontinuity in the v vs. c plot in fig 1 and 2 and the rise in v near saturation is due to anion-anion interaction with water.

## 4. Summary:

The anomalous nature of variation of ultrasonic wave velocity v observed with the increase in concentration c in aqueous solutions of NaCl, KCl and CsCl has been considered for an explanation in view of other experimental studies and simulation results relating structure making/breaking effect of metallic ions in water. It is inferred that $Na^+$, $K^+$, $Cs^+$, and $Cl^-$ ions are in descending order with respect to structure making property. $Na^+$ ions, being strong structure makers, favor the orientation of hydrogen bonded water molecules around it forming more compact network structure and consequent rise in v with c is observed in the lower concentration range in NaCl solution. At relatively higher c, perturbation caused by $Cl^-$ ions induces breaking of HB structure and this is reflected as a discontinuity in the v vs. c plot. $K^+$ ions in the low c range acts as structure breaker of existing HB structure in water and this is reflected as an initial decrease in v with increasing c. With further increase in c, restructuring of water molecules around $K^+$ ions occur and a similar nature of variation of v with c as found in NaCl is observed. For CsCl both of the ions induces structure breaking effect resulting a sharp decrease in v with the increase in c. Near saturation the nature is similar as those observed in NaCl or KCl.